\documentclass[conference]{IEEEtran}
\IEEEoverridecommandlockouts
\usepackage{cite}
\usepackage{amsmath,amssymb,amsfonts}
\usepackage{algorithmic}
\usepackage{graphicx}
\usepackage{textcomp}
\usepackage{xcolor}
\usepackage{booktabs}
\usepackage[hyphens]{url}
\usepackage[hidelinks]{hyperref}

\newif\ifblind\blindfalse

\def\BibTeX{{\rm B\kern-.05em{\sc i\kern-.025em b}\kern-.08em
    T\kern-.1667em\lower.7ex\hbox{E}\kern-.125emX}}

\begin{document}

\title{Cyber-physical WebAssembly: \\ Secure Hardware Interfaces and Pluggable Drivers
\thanks{This work has been partially funded by the European Union (Horizon Europe, Smart Networks and Services Joint Undertaking, ELASTIC project, Grant Agreement number 101139067, https://elasticproject.eu/).\\

\textcopyright~2025 IEEE. Personal use of this material is permitted. Permission from IEEE must be obtained for all other uses, in any current or future media, including reprinting/republishing this material for advertising or promotional purposes, creating new collective works, for resale or redistribution to servers or lists, or reuse of any copyrighted component of this work in other works.}
}

\author{
\IEEEauthorblockN{
\ifblind
Anonymous authors
\else
Michiel Van Kenhove\IEEEauthorrefmark{1},
Maximilian Seidler\IEEEauthorrefmark{2},\\
Friedrich Vandenberghe\IEEEauthorrefmark{1},
Warre Dujardin\IEEEauthorrefmark{1},
Wouter Hennen\IEEEauthorrefmark{1},\\
Arne Vogel\IEEEauthorrefmark{2},
Merlijn Sebrechts\IEEEauthorrefmark{1},
Tom Goethals\IEEEauthorrefmark{1},\\
Filip De Turck\IEEEauthorrefmark{1}
and Bruno Volckaert\IEEEauthorrefmark{1}
}\\
\IEEEauthorblockA{
\begin{minipage}{.5\textwidth}
\centering
\IEEEauthorrefmark{1}IDLab, Department of Information Technology\\
Ghent University - imec, Ghent, Belgium\\
michiel.vankenhove@ugent.be
\end{minipage}%
\begin{minipage}{.5\textwidth}
\centering
\IEEEauthorrefmark{2}System Software Group, Department of Computer Science\\
Friedrich-Alexander-Universit\"at, Erlangen-N\"urnberg, Germany\\
maximilian.seidler@fau.de
\end{minipage}
}
\fi
}

\maketitle

\begin{abstract}
The rapid expansion of Internet of Things (IoT), edge, and embedded devices in the past decade has introduced numerous challenges in terms of security and configuration management. Simultaneously, advances in cloud-native development practices have greatly enhanced the development experience and facilitated quicker updates, thereby enhancing application security. However, applying these advances to IoT, edge, and embedded devices remains a complex task, primarily due to the heterogeneous environments and the need to support devices with extended lifespans. WebAssembly and the WebAssembly System Interface (WASI) has emerged as a promising technology to bridge this gap. As WebAssembly becomes more popular on IoT, edge, and embedded devices, there is a growing demand for hardware interface support in WebAssembly programs. This work presents WASI proposals and proof-of-concept implementations to enable hardware interaction with I2C and USB, which are two commonly used protocols in IoT, directly from WebAssembly applications. This is achieved by running the device drivers within WebAssembly as well. A thorough evaluation of the proof of concepts shows that WASI-USB introduces a minimal overhead of at most 8\% compared to native operating system USB APIs. However, the results show that runtime initialization overhead can be significant in low-latency applications.
\end{abstract}


\section{Introduction}

The Internet of Things (IoT) refers to the network of devices and electronics embedded with various sensors, software, and other technologies to gather and exchange information with other devices and systems over the internet~\cite{leukert_iot_2016}. During the past decade, IoT has experienced remarkable growth in a variety of industries and has become an integral part of our daily lives with more than 17 billion devices connected in 2024~\cite{vailshery_number_2023}.

A large amount of IoT devices are integral to critical infrastructure~\cite{lee_cyber_2008} and operate on the edge of the internet, where the management of their software and firmware poses significant challenges due to their inherent complexity~\cite{gustin_iot_2022}. In accordance with the Cyber Resilience Act~\cite{european_commission_eu_2024}, manufacturers must ensure effective vulnerability management and provide security updates for at least the time the product is expected to be in use. Moreover, the automotive industry is increasingly adopting the practice of wirelessly distributing software updates to vehicles, known as over-the-air updates, where security is of critical importance~\cite{ghosal_secure_2022,oliveira_overair_2024}. In parallel, Industrial Internet of Things systems often feature devices with operational lifespans of more than 30 years~\cite{breivold_internet_2015,stouffer_guide_2023}. To ensure forward compatibility, these systems must be able to integrate with newer hardware components, such as updated sensors and displays, since the original hardware usually reaches end of life when they need to be replaced. This necessitates modularity at the driver level. Additionally, these systems are often tightly integrated with devices that have limited hardware capabilities, increasing the need for minimal overhead.

Although cloud-native development practices have introduced benefits such as isolation, componentization, containerization, and orchestration platforms, these advances are difficult to apply to edge environments and even more so for IoT devices, embedded systems, and microcontrollers~\cite{lwakatare_towards_2016}. Ensuring drivers are secure and up-to-date is crucial for supply chain security, as drivers often have elevated system privileges or direct access to hardware. W3C~\cite{noauthor_w3c_nodate}~WebAssembly (Wasm)~\cite{noauthor_webassembly_nodate} and the WebAssembly System Interface (WASI)~\cite{noauthor_wasi_nodate} emerges as a promising solution to shrink the trusted computing base and enhance embedded development and maintainability through a trusted runtime, modern development methodologies, toolchains and software development kits. Wasm was originally created to execute binary code within the browser to improve the performance of web applications~\cite{haas_bringing_2017}, and is now actively used outside the browser via WASI~\cite{spies_evaluation_2021}. WASI facilitates operating system communication within Wasm by providing a set of portable Application Programming Interfaces (APIs). However, embedded and IoT applications require access to hardware interfaces to enable communication with the system in which they operate. Wasm does not currently meet the demands of cyber-physical systems, particularly in accessing connected hardware. Standardized WASI interfaces are needed in Wasm runtimes to ensure portability and enable communication with hardware across diverse devices.

In this work, we propose standardized WASI interfaces and proof-of-concept implementations for connecting hardware using I2C and USB, along with security mediation and isolation, on low-powered embedded and IoT devices running a Wasm runtime with pluggable drivers. Our solution aims to enhance supply chain security by reducing the number of components that need to be trusted, addressing the inherent insecurity of native drivers. Furthermore, we emphasize the importance of forward compatibility and modularity at the driver level to ensure adaptability and security in evolving technological landscapes. Specifically, this paper aims to answer the following \textsc{\textbf{Research Questions}}:

\textit{\textbf{RQ 1:}} How to enable WebAssembly applications outside the browser to interact with hardware interfaces such as I2C and USB?

\textit{\textbf{RQ 2:}} How to apply access control security measures in WebAssembly to ensure authorized access to hardware interfaces?



\textit{\textbf{RQ 3:}} How does our solution to interact with hardware interfaces through WebAssembly impact performance compared to native solutions?


We begin by positioning our work within the existing literature in Section~\ref{sec:related_work}. Then, we provide a threefold contribution to the state of the art. Specifically, our \textbf{\textsc{Contributions}} are:

\textit{\textbf{Contribution 1:}} Develop a novel solution to enable Web\-As\-sem\-bly-based workloads on IoT and edge devices to interact with (I2C and USB) hardware interfaces (Section~\ref{sec:architecture}). Specifically, our solution tackles the challenge of giving sandboxed WebAssembly applications mediated access to device resources.

\textit{\textbf{Contribution 2:}} Demonstrate the feasibility of our solution through the development of multiple proof-of-concept implementations (Section~\ref{sec:implementation}).

\textit{\textbf{Contribution 3:}} Evaluate our solution by conducting benchmarks and measuring performance overhead compared to native hardware access (Section~\ref{sec:evaluation}).



\section{Related Work}\label{sec:related_work}

WebAssembly has emerged as a promising technology for enabling high-performance, cross-platform applications, particularly in the context of embedded systems and IoT devices. Several studies have explored the integration of Wasm with various hardware and software environments to enhance the capabilities and security of these systems, which are discussed below.

Singh et al.~\cite{singhWARDuinoDynamicWebAssembly2019} developed the WARDuino virtual machine (VM), which exposes common IoT Arduino APIs to the Wasm runtime. WARDuino only facilitates the use of Wasm in, as its name suggests, Arduino-based projects. This poses notable limitations in terms of portability. This also means that the hosting device must provide all the necessary APIs and that there is no hardware support beyond the Arduino ecosystem.

Aerogel~\cite{liuAerogelLightweightAccess2021} is a lightweight access control framework that addresses security gaps between the bare-metal IoT devices and the Wasm execution environment concerning access control for sensors, actuators, processor energy usage, and memory usage. Aerogel enhances the security and provides resource management in multi-tenant Wasm runtime environments, where each Wasm application is considered a tenant. A user-defined access control specification sheet allows one to set restrictions on specific inputs/outputs (I/Os), limitations on the number of accesses within a specified time frame, an upper bound on memory and CPU usage, and the maximum number of applications having shared access to a specific hardware component. Access to the requested resource will only be granted if all access control conditions are met.

WiProg~\cite{liWiProgWebAssemblybasedApproach2021} is a framework that simplifies integrated IoT programming and application development for sensor-to-edge systems using Wasm, focusing on application offloading. It allows for a monolithic application development approach, enabling offloading based on compiler annotations and policies. WiProg provides hardware-agnostic APIs for analog I/O, digital I/O, and Universal Asynchronous Receiver-Transmitter (UART), but lacks support for important protocols like I2C.

Li et al.~\cite{liBringingWebassemblyResourceconstrained2022} proposed WAIT, a lightweight Wasm runtime for IoT devices in device-cloud integrated applications. It enables Wasm ahead-of-time (AOT) compilation on resource-constrained devices. Upon receiving a Wasm module binary, WAIT performs on-device AOT compilation, checks sandboxing guarantees, and executes the module. Native APIs for peripherals and I/O are exposed via the Wasm \texttt{import} mechanism. The AOT compiler then uses binary rewriting to replace imported function calls with direct assembly code, reducing external function call overhead.

Finally, Wasmachine~\cite{wenWasmachineBringEdge2020} is a bare-metal operating system for embedded devices, running everything in kernel mode for zero-cost system calls. It exports system calls in WASI prototypes and runs each Wasm application as a kernel thread, allowing the application to invoke system calls as normal function calls without additional context switching costs. Hardware access is provided using WASI with per-application permissions. Despite its innovative approach, it has certain limitations, including limited CPU architecture support and the reliance on a Unix-like monolithic kernel architecture, which may affect its feasibility on some IoT or edge devices.

\section{Architecture}\label{sec:architecture}

\begin{figure*}[ht]
    \centering
    \includegraphics[width=0.8\textwidth]{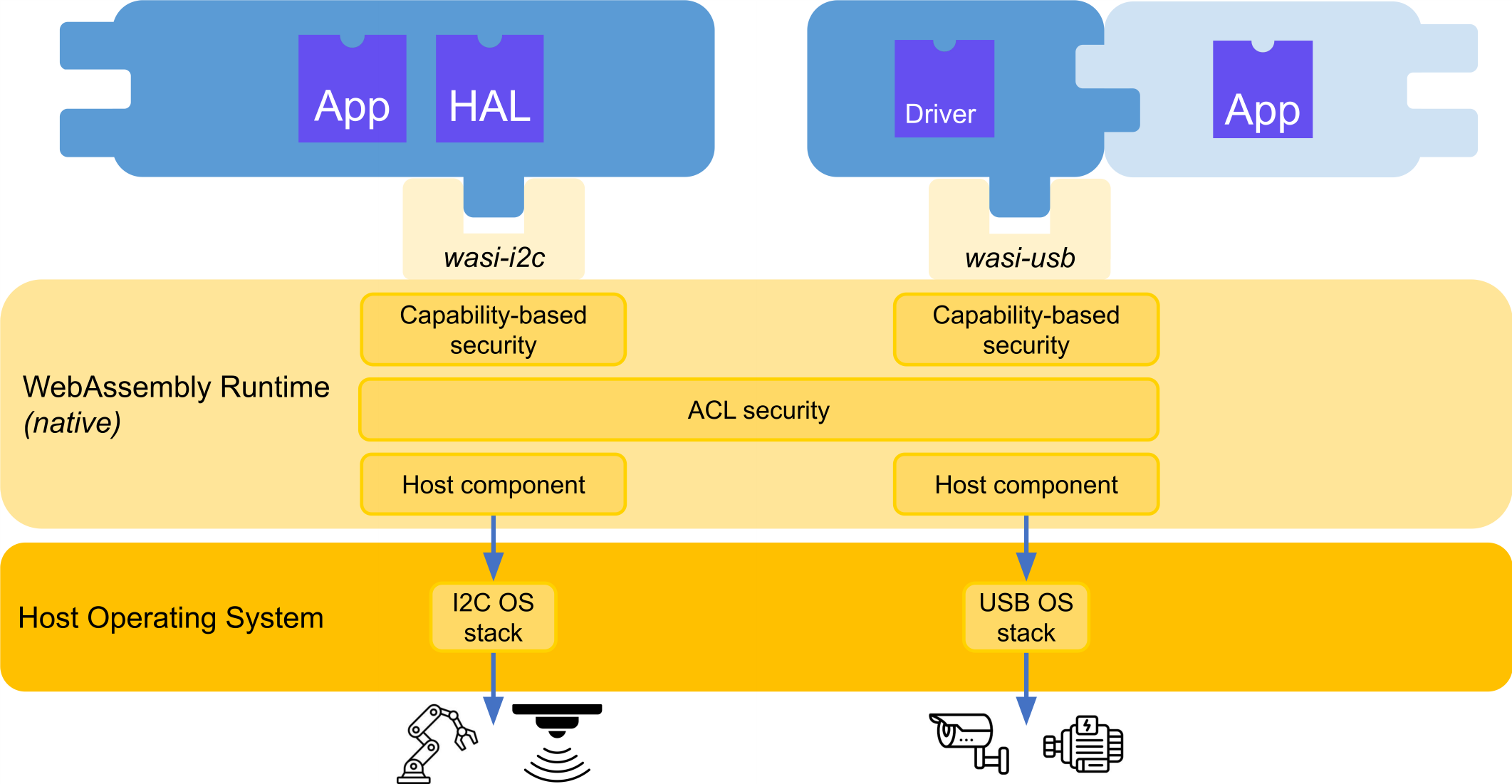}
    \caption{Schematic representation of the proof-of-concept architecture that enables I2C and USB hardware interface support in Wasm applications. The application and device drivers run as Wasm guest components and connect to the host components via WASI. The host components provide ACL and capability-based security and use underlying OS-specific APIs to implement the interfaces. This figure contains resources from Flaticon.com.}\label{fig:architecture_diagram}
\end{figure*}

The proof-of-concept architecture designed by the authors to enable mediated (I2C and USB) hardware interface support in Wasm applications is shown in Fig.~\ref{fig:architecture_diagram}. Both the application and the peripheral device drivers run as \textbf{WebAssembly components}. These connect to the runtime via standardized generic I2C and USB \textbf{WebAssembly System Interfaces}. The runtime itself has a number of \textbf{host components}, one for each interface, that implement the interfaces and provide access-control list (\textbf{ACL}) and \textbf{capability-based security}. The host component itself communicates with the host operating system (OS) and uses \textbf{OS-specific APIs} for hardware access. The remainder of this section examines important concepts of our novel solution in more detail.

\subsection{Interface Worlds}

An interface world describes the system calls that are supported by the runtime using the Component Model's Interface Description Language (IDL) called Wasm Interface Type (WIT). For example, the USB world contains a WIT description of all the system calls that Wasm components can use to communicate with USB devices.

This world is a high-level interface contract between the runtime and the guest component and can be used by tools such as \texttt{wit-bindgen}~\cite{noauthor_bytecodealliancewit-bindgen_2024} to automatically generate bindings for calling these APIs from various programming languages. The source code of the guest components uses these bindings to implement the world and to call the imported APIs, allowing them to interact with hardware devices supported by the interface. The runtime host component can use these bindings to implement and export the functions that make up the API.

\subsection{Host Component}

The host component is the host-side implementation of the system call API. It is embedded in the runtime and runs as native code directly on the underlying OS, as it requires access to native APIs of the host OS to communicate with hardware devices. As such, the host component is highly dependent on the specific host OS interfaces. Because it is embedded in the runtime, the host component also depends on the specific implementation details of the runtime. Therefore, the implementation of the host component is unique for every combination of runtime and underlying (operating) system. Any new system or runtime that wishes to support certain worlds, will require a modified implementation of the host component. However, this burden can be lessened by using higher-level cross-platform libraries to implement the host component, rather than depending directly on the OS through system calls. This way, one implementation of the host component can be compiled for multiple operating systems. Unfortunately, using a higher-level library is not always feasible, especially on certain low-powered embedded devices due to device-specific limitations, requiring a custom host component implementation for those devices.

\subsection{Guest Components}

Guest components contain the application logic and device drivers or a hardware abstraction layer (HAL). These can be made up of multiple separate components, or the application logic can embed the driver and HAL logic inside of it. A guest component specifies which interfaces it requires through imports and the interfaces it provides through exports using a WIT world description. This informs the runtime about which host component interfaces to connect to the guest component, allowing the guest component to interact with hardware devices through host component function calls. Guest components are contained within a sandbox by the runtime, ensuring they remain isolated and cannot access system resources without authorization.

\subsection{Runtime}

The Wasm runtime resolves the imports and exports for all components and links them together. To successfully resolve imports and exports, the runtime must support the WIT world that a component implements. This can be achieved by a host component providing the necessary implementations for the guest component's imported functions. After linking, the guest component has access to the imported functions from the world it implements. The runtime, in turn, has access to the functions that the world exports. To begin execution, the runtime must call one of those exported functions. Since interface worlds are not concerned with how a component starts execution, this is delegated to the \texttt{wasi-cli}~\cite{webassembly_webassemblywasi-cli_2024} world to specify the entry-point. Popular Wasm runtimes include Wasmtime~\cite{noauthor_wasmtime_nodate} and WebAssembly Micro Runtime (WAMR)~\cite{bytecode_alliance_wamr_webassembly_nodate}, which are both being developed by the Bytecode Alliance~\cite{noauthor_bytecode_nodate}.

\subsection{Capability-Based Security}\label{section:capability_based_security}

One of the strengths of WASI is its capability-based security mechanism. This mechanism implicitly allows granular control over what resources an application can access based on the API definition. If a resource is not provided, the module cannot interact with it, implicitly revoking the capability for that resource.

Although WASI enables the implementation of capability-based security for system interfaces, it does not stipulate the level of granularity at which these security controls should be constructed. To illustrate, a USB API may employ capability-based access control at multiple levels:

\begin{itemize}
\item Device level: Allow only devices with specific vendor and product ID pairs.
\item Configuration level: Allow only devices with specific configuration types, e.g., self-powered devices.
\item Interface level: Allow only interfaces with specific property types, e.g., those with the mass storage class code.
\item Endpoint level: Allow only endpoints with specific property types, e.g., those with an \texttt{IN} direction that restricts the guest to receiving data only.
\end{itemize}

The capability-based security model is inherently defined by the API, which also determines the granularity of the security measures. This implies that access control must define the resources to be protected, which is impractical for fine granularity and challenging when control layers are orthogonal or complementary. Consequently, we propose an extension of capability-based security with explicit access control lists.

\subsection{ACL-Based Security}

Access control lists add a layer of granular and orthogonal access control, complementing capability-based access control. Unlike capability-based access control, which requires changes to the API, ACLs can be implemented without affecting the existing API. For example, an ACL rule might allow an application to read from a device but not write to it. Furthermore, ACLs offer greater flexibility and can be fine-tuned to the environment and runtime as needed. Based on these advantages, we propose that standardized interfaces use capability-based security for high-level concepts and use an additional ACL to further restrict permissions.

Our implementation of ACL-based access control necessitates the incorporation of logic specific to the runtime, environment, and peripheral. Whenever the runtime creates a new module instance, an I/O context is created and linked to the corresponding execution environment. Whenever a device interaction is performed, the corresponding I/O context is retrieved from the execution environment, and depending on the type of device interaction and the I/O context's properties, the action is permitted or denied.

The I/O context's properties may refer to permitted interaction types, such as read or write. Additionally, specific values may be blocked, which is necessary for device interaction with arbitrary data, but, for example, prevents changes in the configuration. Furthermore, the context may specify how connections should be established with end devices, preventing communication with other peripherals sharing the same bus.

\section{Implementation Details}\label{sec:implementation}

To validate our proposed cyber-physical WebAssembly architecture, we developed three proof-of-concept implementations, each addressing different aspects of the solution. Our primary objectives were to confirm the feasibility of the approach and identify potential opportunities or shortcomings in the underlying architecture and standardization methodologies.


\begin{itemize}
    \item \textbf{I2C in Wasmtime}: Demonstrates driving an I2C seven-segment display from Wasm on a Linux host and a Raspberry Pi Pico, and reading temperature and humidity data from an HTS221 sensor on a Raspberry Pi 3B.
    \item \textbf{I2C in WAMR}: Demonstrates reading data from an HTS221 sensor. Due to WAMR's current limitations, including the lack of WASI Preview 2 support, this solution uses an ACL-based security system instead of the component model or capability-based security.
    \item \textbf{USB in Wasmtime}: Demonstrates Wasm applications communicating with USB devices, including a flash drive and an Xbox controller using Wasm-based drivers.
\end{itemize}


\subsection{I2C in Wasmtime}

The I2C in Wasmtime proof-of-concept implementation~\cite{noauthor_idlab-discoveri2c-wasm-components_2024} is made up of several components: the host implementation, two WIT worlds and two guest implementations.

The host implementation handles the I2C connection with the device, whereas the guests are responsible for device-specific logic, such as reading temperature or displaying time. The host and guest components interact through two WIT worlds, one for the display and the other for the HTS221 sensor. These WIT worlds import the necessary interfaces from the WASI-I2C~\cite{webassembly_webassemblywasi-i2c_2024} proposal that are implemented by the host, allowing the guest to communicate with the I2C device. Additionally, the WIT worlds export device-specific functions implemented by the guest. Although these worlds were written for our specific proof of concepts, they aim to be as generic as possible to allow future support for other displays and sensors.

The guest implementations do not use Rust's \texttt{std} library, only importing necessary types individually, which helps minimize the size of the compiled Wasm binaries. The guest component for the seven-segment display directly writes bytes to control it, while the HTS221 sensor component leverages the \texttt{embedded-hal}~\cite{noauthor_rust-embeddedembedded-hal_2024} compatible \texttt{hts221}~\cite{friedrich_zelzahnhts221_2024} Rust crate for simplicity. To adapt the \texttt{embedded-hal} API to the slightly different generated bindings, the \texttt{wasi-embedded-hal}~\cite{noauthor_idlab-discoverwasi-embedded-hal_2024} crate was developed.

\subsection{I2C in WAMR}

WAMR currently does not support WASI Preview 2. Consequently, the component model and WIT are not available, which means that a standardized I2C connection cannot be passed to the guest. As a workaround, the I2C connection remains global on the host side. For the host to bind with the guest, it has to register the function that the guest imports and find the functions that the guest exports.


\subsection{USB in Wasmtime}

The proof-of-concept implementation~\cite{noauthor_idlab-discoverusb-wasm_2024,webassembly_webassemblywasi-usb_2024} of USB in Wasmtime includes a host implementation along with several guest implementations. The main concept of the USB implementation in Wasmtime mirrors that of the I2C implementation in Wasmtime: the host component is platform dependent and interacts with the underlying system to access the USB resource, while the guest components handle device-specific logic, such as reading the inputs of an Xbox controller.

To facilitate the development of the host component and enhance its platform independence, \texttt{rusb}~\cite{averyanov_a1ienrusb_2024}, a Rust wrapper around \texttt{libusb}~\cite{noauthor_libusblibusb_2024}, is used. The USB flash drive guest component supports various USB Attached SCSI device commands, including basic file system operations like listing directory contents and reading files. The file system support is fully integrated within the guest component itself using the \texttt{fatfs}~\cite{harabien_rafalhrust-fatfs_2024} Rust crate, without relying on the host operating system's file system capabilities. The Xbox guest component continuously monitors the current state of the controller through its USB interrupt pipe and writes button presses and joystick positions to standard output.

\section{Evaluation Results}\label{sec:evaluation}

This section assesses the practicality of our proposals in real-world scenarios. Initially, each proof of concept is evaluated to ensure it meets the functional requirements, typically by manually verifying device functionality. Subsequently, various benchmarks are performed to identify any performance implications.

\subsection{I2C Performance Evaluation}

The I2C proof of concept for the seven-segment display is evaluated on a Raspberry Pi 4B, whereas the HTS221 sensor proof of concept is evaluated on a Raspberry Pi 3B. The display evaluation will compare the native, Wasmtime, and WAMR implementations, while the HTS221 sensor evaluation will compare the native and Wasmtime implementations. All benchmark results are available~\cite{noauthor_idlab-discoveri2c-wasm-components_2024} on GitHub.

\begin{figure}[tbp]
    \centerline{\includegraphics[width=1\linewidth]{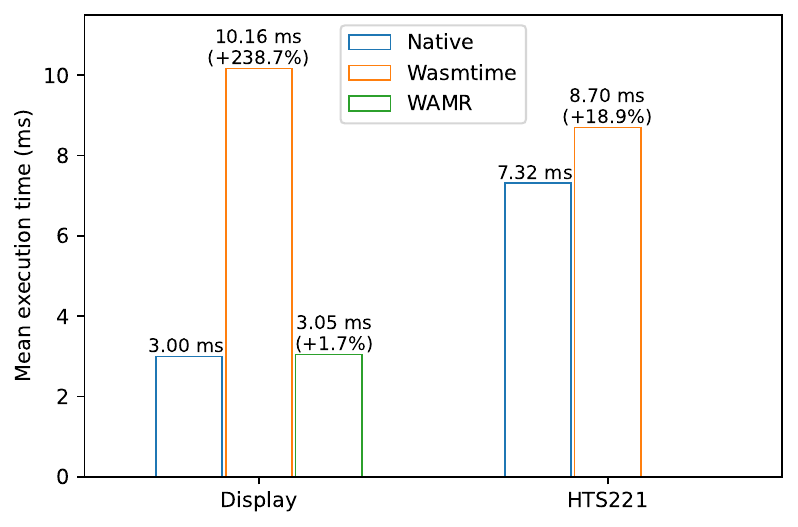}}
    \caption{Mean cold start execution times for writing a four-digit number to the display and reading the temperature from the HTS221 sensor, comparing native, Wasmtime, and WAMR implementations.}\label{fig:i2c_exectime_bar}
\end{figure}

To assess performance overhead, we conduct 100 iterations of writing a four-digit number to the display and also reading the temperature from the HTS221 sensor. To highlight the worst-case overhead, cold start execution times are measured, where each measurement includes the time required to instantiate the Wasm binary within the runtime. The mean cold start execution times are then compared across the native, Wasmtime, and WAMR implementations. Fig.~\ref{fig:i2c_exectime_bar} illustrates the mean cold start execution times for the two proof of concepts. Wasmtime shows a notable overhead relative to the native implementation, with the mean cold start execution times being 7.16~ms (+238.7\%) longer for the display and 1.38~ms (+18.9\%) longer for the HTS221 proof of concept. In contrast, the WAMR implementation incurs only a minimal overhead of 50~\(\mu\)s (+1.7\%). Most of the additional time consumed by Wasmtime can be attributed to functions associated with Cranelift~\cite{noauthor_cranelift_nodate}, a compiler backend utilized by Wasmtime for both just-in-time and ahead-of-time compilation.

\subsection{USB Performance Evaluation}

The USB proof of concept is evaluated on three systems: one with an AMD Ryzen 7 5800H and USB3.0 running x86 Linux 6.8.9, a Raspberry Pi 3B+ with USB2.0 running AArch64 Linux 6.8.0, and a third with an AMD Ryzen 7 5800X3D and USB3.0 running x86 Windows 10 22H2. All benchmark results are available~\cite{noauthor_idlab-discoverwasi-usb-benchmarks_2024} on GitHub.

To assess the runtime size and memory usage overhead introduced by the USB host component, we compare the Wasmtime runtime both with and without the USB host component. Our measurements show that the runtime size overhead is minimal, adding only 160~KiB (+1.1\%) for the x86 Linux runtime, 220~KiB (+1.7\%) for the AArch64 Linux runtime, and 275~KiB (+1.9\%) for the Windows runtime. Memory usage overhead is also minimal for the Linux runtimes, comprising only 459~KiB (+1.7\%) for the x86 Linux runtime and 295~KiB (+1.5\%) for the AArch64 Linux runtime. The Windows runtime incurs a more severe memory overhead of 1427~KiB (+11.0\%), which can be attributed to an entirely different backend being used by \texttt{libusb} on Windows.

\begin{figure}[tbp]
    \centerline{\includegraphics[width=1\linewidth]{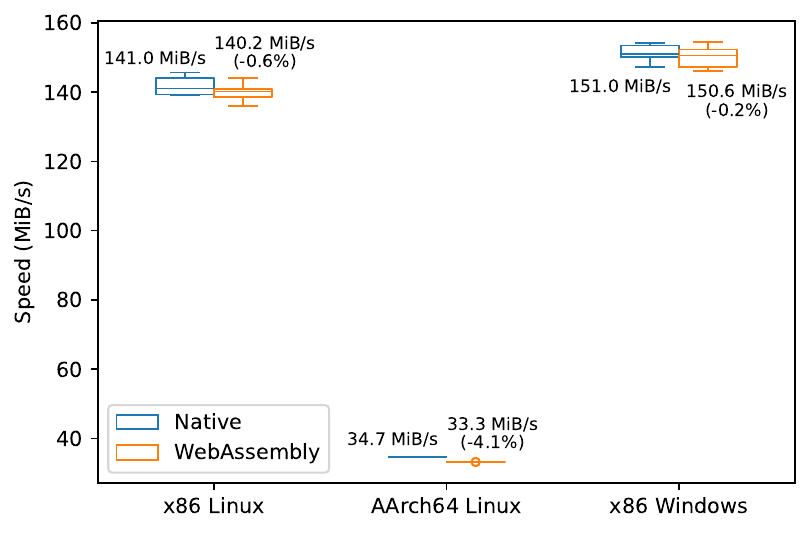}}
    \caption{USB flash drive sequential read speeds across various platforms for both native and Wasm guest applications show minimal overhead on high-powered machines, while lower-powered devices experience a larger overhead. The reported speeds represent the median throughput.}\label{fig:seq_read_boxplot}
\end{figure}

The impact on throughput is assessed using a USB flash drive to read or write as many data blocks as quickly as possible. Fig.~\ref{fig:seq_read_boxplot} depicts the results for a sequential read benchmark. On x86 Linux and Windows, the overhead is minimal, respectively resulting in only a 0.8~MiB/s (-0.6\%) and 0.4~MiB/s (-0.2\%) reduction in median throughput. The overall throughput on AArch64 Linux is notably lower, along with a 1.4~MiB/s (-4.1\%) decrease in median throughput in Wasm. This can be attributed to the fact that the Raspberry Pi 3B+ only supports USB2.0, and this also implies that the overhead is larger on less powerful hardware. Despite this, the overhead remains relatively modest. Benchmarks for sequential write, random read, and random write show similar relative overheads.

\begin{figure}[tbp]
    \centerline{\includegraphics[width=1\linewidth]{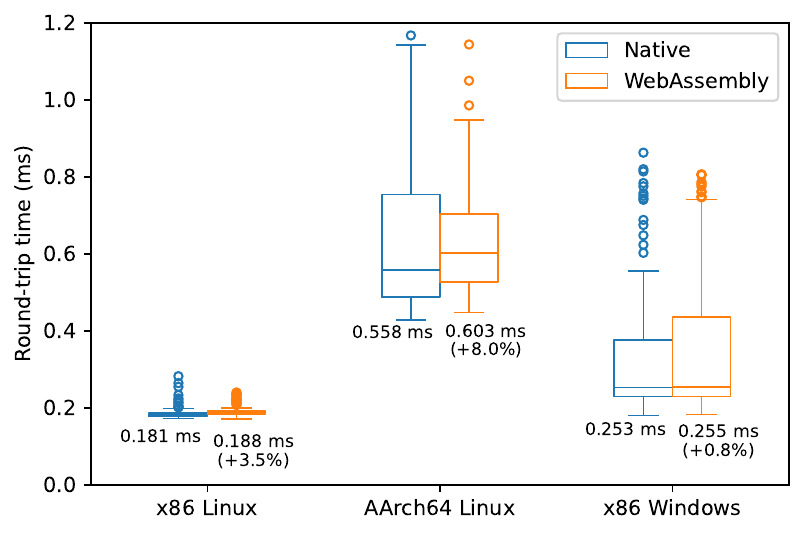}}
    \caption{Round-trip time for a 32-byte USB bulk transfer across different platforms, comparing native and Wasm guest applications. x86 Linux shows minimal outliers and reasonable overhead, whereas AArch64 displays greater overhead. The reported times represent the median round-trip time.}\label{fig:rtt_bulk_boxplot}
\end{figure}

The latency overhead is evaluated using an Arduino Nano 33 IoT device with custom firmware that receives a message from the host device and echoes it back to the host without modifications. This allows the host to measure the round-trip time (RTT) of this message. Fig.~\ref{fig:rtt_bulk_boxplot} shows the observed RTTs for a bulk transfer of a 32-byte message. Due to the low priority of bulk transfers among USB transfer modes, numerous outliers are present. Some outliers have been excluded to facilitate the comparison of the median RTTs. The overhead introduced by Wasm is relatively modest. On x86 Linux, it results in a median RTT that is only 7~\(\mu\)s (+3.5\%) longer than the native implementation. On x86 Windows, an even lower overhead of 2~\(\mu\)s (+0.8\%) is observed. However, on AArch64, the overhead is more pronounced, with an increase of 45~\(\mu\)s (+8.0\%).

\section{Conclusion}\label{sec:conclusion}

With the growing popularity of WebAssembly on embedded and IoT devices, the need for standardized hardware interface support in WebAssembly arose. This paper introduces two WebAssembly System Interface proposals, WASI-I2C and WASI-USB, enabling the use of I2C and USB within WebAssembly. Our cyber-physical WebAssembly proof-of-concept architecture is described in depth, including a solution for authorized hardware interface access. Hardware interface access is managed through the capability-based security mechanism of WASI, and is extended with explicit ACL-based security for more granular access control. Proof-of-concept implementations showcase the feasibility of the proposed architecture and are evaluated to confirm that they meet functional requirements and to identify any performance impacts. Our evaluation results show that the overhead introduced by WASI-USB is minimal, resulting in a maximum throughput reduction of 1.4~MiB/s (-4.1\%) and an increase in round-trip time of up to 45~\(\mu\)s (+8.0\%). Our test results for WASI-I2C include the initialization of the Wasm binary within the runtime in each measurement. This reveals significantly worse overhead in Wasmtime, with the mean cold start execution time increasing by up to 7.16~ms (+238.7\%) at worst and 1.38~ms (+18.9\%) at best. This emphasizes that the overhead of runtime and binary initialization is significant, indicating that low-latency applications require a warm-up period.

This work has established a foundation for standardizing the I2C and USB WASI interfaces, and demonstrates the feasibility of a cyber-physical WebAssembly future on IoT and embedded devices. However, future work will be necessary to stabilize the standardization proposals, which will involve extensive testing, implementation across multiple runtimes, and gathering community feedback. Additionally, these proposals lack asynchronous support, which limits their functionality in certain environments.

\section*{Acknowledgment}

We extend our heartfelt gratitude to everyone who contributed to these standardization proposals: Valentin Olpp, Dan Gohman, Emiel Van Severen and the Bytecode Alliance \& WASI subgroup of the W3C WebAssembly Community Group. This work received funding from the Smart Networks and Services Joint Undertaking (SNS JU) under the Horizon Europe research and innovation program of the European Union under Grant Agreement number 101139067 (https://elasticproject.eu/). Expressed views and opinions are however those of the authors only and do not necessarily reflect those of the European Union. Neither the European Union nor the granting authority can be held responsible for them.

\bibliographystyle{IEEEtran}
\bibliography{references}

\end{document}